# Efficient Video Indexing on the Web: A System that Leverages User Interactions with a Video Player

Ioannis Leftheriotis[1], Chrysoula Gkonela[1] and Konstantinos Chorianopoulos[1]

[1]Ionian Univesity, Department of Informatics, 7 Tsirigoti square, 49100 Corfu, Greece
{c09levth, c09gkon, choko}@ionio.gr

**Abstract.** In this paper, we propose a user-based video-indexing method, that automatically generates thumbnails of the most important scenes of an online video stream, by analyzing users' interactions with a web video player. As a test bench to verify our idea we have extended the YouTube video player into the VideoSkip system. In addition, VideoSkip uses a web-database (Google Application Engine) to keep a record of some important parameters, such as the timing of basic user actions (play, pause, skip). Moreover, we implemented an algorithm that selects representative thumbnails. Finally, we populated the system with data from an experiment with nine users. We found that the VideoSkip system indexes video content by leveraging implicit users interactions, such as pause and thirty seconds skip. Our early findings point toward improvements of the web video player and its thumbnail generation technique. The VideSkip system could compliment content-based algorithms, in order to achieve efficient video-indexing in difficult videos, such as lectures or sports.

## 1 Introduction

During the last years, the intense growth of the internet has given impetus to sharing of video material between users all over the world. Nowadays, User Generated Content together with movies, tv series, lectures, sports video, news etc., are all available to the public. One of the most successful platforms that millions of users use on a daily basis in order to browse all these videos is YouTube.

Users, apart from watching videos on the main YouTube video player, can upload videos or perform other important tasks, such as commenting videos, replying with other videos, tagging or publishing videos in other platforms. A lot of research has been done so as to prove the importance of the comments and other metadata information for complex processes, such as generating important thumbnails of the video or even producing automatically generated summaries. But, although there is a variety of methods that collect and manipulate all these information, the majority of them is usually burdensome for the users. Moreover, the percentage of users leaving a comment is too small according to the real number of viewers of a video.

In this work, we suggest the idea that the best way to extract useful information about a video, is to simply let the viewer browse the video as if he was in the YouTube web player, and just store all the interactions with the player (e.g. play, pause) for future use. As Shamma et al[8] proved, the more the emotive energy of a scene, the more the specific interval of the video containing that scene is used. Combining that result with the fact that we are proposing to record all the interactions between the user and the web player in order to infer the most important scenes of a video, and to automatically generate thumbnails, or even implement a summarization feature.

Bearing this perspective in mind, we implemented a web video player based on the YouTube API (application programming interface) and we linked it with a small database by using Google APP Engine infrastructure. In the remaining of the paper, apart from presenting the web video player, we examine the properties that this player should have, discussed some considerations about the video content of the player, and finally, we present the results of a pilot usability test we conducted with 9 subjects.

## 2   Related Work

A lot of research has been done in order to improve users' browsing experience while watching video content. Some researchers use automated summarization techniques. Takahashi et al[10] for example, propose a summarization method for sports video that uses metadata to extract the important video segments. Sports video are being processed beforehand and information such as play ranks, play occurrence time and number of replays are used to generate a video summary whose duration is selected by the user.

Another technique for improving browsing experience is using methods that generate thumbnails. For example, SmartSkip[1] from Microsoft Research is an interface that uses the histogram of images in almost every 10 seconds of the video and looking at rapid overall changes in the color and brightness, generates thumbnails. Li, Gupta et al[5]developed an interface that generates shot boundaries using a detection algorithm that identifies transitions between shots. Although the table of contents of the video is pre-generated, they use techniques such as time compression (increase playback speed) and pause removal (detects and removes pauses and silence segments of the video) that help the user browse the video more efficiently. These approaches are not complete, because they are content-based. Even though they generate important content for the user they do not take into account his preferences. In this research, we investigate the indexing of video content with user-based methods.

On the other hand, researchers have realized that the viewer is not the end of the video production – distribution – consumption chain. Viewer is capable of being a significant node in the chain, playing different roles such as distributor or even producer of the content. In particular, user interactions with the video add value to the content. The cumulative user interactions could be leveraged for the benefit of future viewers. For example, some browsing approaches focus on personalization with the user, such as the application framework proposed by Hjelsvolt et al[3] from Siemens, which proposes hotspots and hyperlinks using a personalized model that matches the content against the user profile. Although, this framework is based on users' preferences, it constantly needs their response to various queries in order to build their profile.

According to Money et al (2008) [7], there are two video summarization techniques, internal summarization which analyze internal information from the video stream produced during the production stage of the video lifecycle (such as in [10]) and external summarization techniques which analyzes external information during any stage of the video lifecycle (such as in [3],[6]). We propose that the same classification applies to thumbnails generation from videos. SmartSkip[1] is a good example of internal thumbnail generation technique. On the other hand, an external one is proposed on [3].

Moreover, in case of external summarization, two types of information are important [7]. User based information, which is information derived from users' behaviors and interactions with the video, and contextual information, which incorporates information from the environment and not directly from the user. For example, Money et al (2009) [6] propose a video summarization technique that takes into account users' physiological response measures such as heart rate or blood volume to indicate memorable or emotionally engaging video content. On the other hand, Gamhewage et al[9] present a system for video summarization in ubiquitous environment that uses pressure-based floor sensors, a characteristic example of contextual information system. Even though these external summarization techniques take into consideration users' interactions and preferences, they seem to be cumbersome for the users and too complicated to be implemented.

In this paper, we propose an external thumbnail generation method based on user interactions. For this purpose, we have developed a web video player, VideoSkip, which collects interactions of users while they watch the video. Unlike other approaches, not only have the users to response queries during the procedure, but they are not aware of this mining process.

## 3  VideoSkip Web Video Player

VideoSkip is a web video player we developed to gather interactions of the users while they watch a video. Based on these interactions, representative thumbnails of the video are generated. In fig.1 we can see a screenshot of the player, presenting the main video player window including various buttons and three proposed thumbnail images below.

### 3.1  Software Tools

In order to implement this web video player, we used a number of software tools such as Google App Engine and YouTube API.

Google App Engine enables building and hosting on the same systems that power Google applications. As a result, users of VideoSkip should have a Google account in order to sign in and watch our videos. Each time a user was logged in the VideoSkip web player, every interaction he had by pressing any button, was recorded and stored in a database provided by Google App Engine for future use (e.g. generating thumbnails).

YouTube API allows developers to use the infrastructure of YouTube and therefore the vast amount of YouTube videos. This API provides us with a chromeless player (a player without controls). We use JavaScript to create our own buttons and implement their functions.

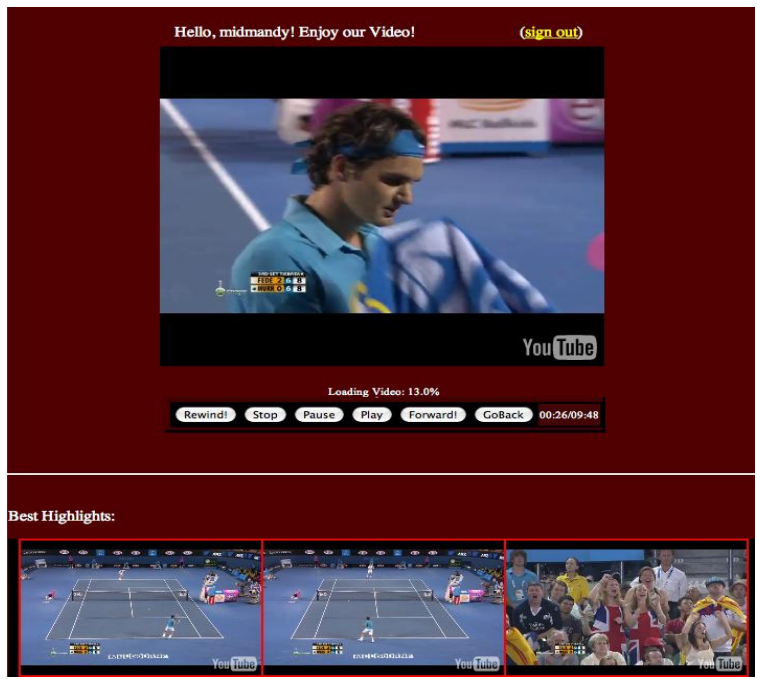

**Fig. 1.** VideoSkip window with the main video player and three proposed thumbnails below

### 3.2 Player Buttons

In fig.1 we can see the buttons of VideoSkip player. Rewind button shows the video on the reverse order by jumping three seconds backwards every half a second. The Rewind button changes its label to inform the user about the reverse state of the video player and the user, in order to stop it, has to push again the same button (Stop Rewind) or the Play button. Stop button simply interrupts the playing state of video, returning the player to its initial state of 00:00 sec. Pause button temporarily freeze the video. Play button starts the video. Forward button shows the video on a faster pace. Video player jumps the next three seconds of the video every half a second. During Rewind and Forward state of the video player, the sound is muted. We decided to use these buttons because they are usually seen in VCR devices and in software playback applications [5] and as a result users should be familiar with them.

In our effort to uncover the favorite scenes of the video considering users' preferences, a new button was added, called GoBack. Its main purpose is to replay the last viewed thirty seconds of the video.

Next to the player's button the current time of the video is shown followed by the total time of the video in seconds.

### 3.3 User Logging

Google App Engine database (the datastore), is used to store users' interactions. Each time a user signs in the web video player application, a new record is created. As it is presented in fig.2, this record includes four fields: a unique id, the username of the user's Google account, the date and a Text variable including all the interactions with the buttons of the web video player.

| ID/Name | author | content | date |
|---|---|---|---|
| id=1 | videoskiptest | play:0.08 fast:9.567 play:44.284 fast:49.11 play:97.963 fast:109.92 play:121.012 replay:130.728 replay:106.255 | 2010-04-28 15:27:00.476000 |
| id=3001 | nikxalkias | play:0.08 fast:37.384 play:48.112 pause:49.459 stop:49.459 rew:0 pause:0 play:0 | 2010-04-28 19:08:25.618000 |

**Fig. 2.** A screenshot of the records' table showing the id, the username (author), the content of the interactions and the date

Whenever a button is pressed, an abbreviation of the button's name and the time it occurred are added to the Text variable. Apparently, the time that is stored is the specific second of the video. The content of the Text variable is used to understand the important segments of the video and therefore generate thumbnails as it is described in the next section.

In order to complete the storing process user must press the Submit and Exit button below the main web video player buttons. When this button is pressed the new record is saved to the datastore, including the exact time of this event. In case of ignoring the Submit and Exit button, the system informs the user with a message box that he has to remain to the web page in order to press that button, before escaping the web video player.

### 3.4 Thumbnail generation algorithm

Every single record is used to generate thumbnails. For each record in the datastore we search for specific button interactions in the content of its Text variable field. Any button or combination could be used.

In our case, taking into account the functionality of GoBack button, we built an algorithm to extract highlight thumbnails. We consider that every video is associated with an array of k cells, where k is the number of the duration of the video in seconds. Initially, the array is empty. Each time user presses the GoBack button the cells' values, matching the last thirty seconds of the video, are incremented by one.

As the number of users who watch the video increases, higher values in specific cells of the array are accumulated. By using a simple sorting technique, the three greatest values of the array are extracted. In order to avoid having consecutive cells as a result, we defined a distance threshold of thirty seconds between them.

The positions of the three greatest values of the array correspond to the time of the most replayed and therefore the most popular video scenes. These three specific scenes are used as proposed thumbnails exactly underneath the web video player.

Whenever a user moves his mouse over a thumbnail, this turns it into a small video player that shows the video from the corresponding time; and in the case of a thumbnail being clicked by the user, the main web video player starts showing the video from the corresponding time respectively.

## 4. Choosing The Video Content

VideoSkip, being a player based on YouTube API, can support a big and growing variety of video content. But, even though large numbers of almost every kind of video are available online, we had to restrict our choices in a small number of videos. Our primary restriction was the use of videos that are as much unstructured as possible, because, the more unstructured a video is, the more important the thumbnail generation for the future viewers will be. Considering the fact that we use users' interactions to generate thumbnails, another important factor is the attractiveness of a video. In order to have as many users as possible, and therefore maximize the number of interactions, we have to use videos that seem to be important, entertaining and enjoyable in general. Another key factor is the time of a video. In general, YouTube allows video uploading limited to 10 minutes. Although we were able to use videos that exceeded that small limit, we decided not to, because we supposed that it would be tiresome for the majority users.

In [5], Li et al, in order to test their enhanced browser user interface, used six different types of video, which were classified in three categories: informational audio-centric videos like classroom lectures and conference presentations, informational video-centric like travel and sports videos and narrative-entertainment like television dramas. It was observed that users tent to have content-specific browsing behaviour. In an effort to evaluate our player we included all three categories. In the next chapter we present the usability test of VideoSkip and the interesting results that occurred.

# 5. Usability Test

Usability is not just the appearance of the user interface. It is related with the way the system reacts with the user and its basic attributes are learnability, efficiency, sufficiency and satisfaction. During the development of any type of project, the steps of usability process should be followed. These steps help user interaction designers to answer crucial questions, during the analysis phase, and supports the design phase. The main tool for this procedure is usability testing that might reveal possible problems[2]. For this purpose there is a variety of tests. A common used usability test that could be applied to VideoSkip is SUS (System Usability Scale) [4].

### 5.1 Methodology

We decided to select 9 participants (of both genders) as the subjects of our test. The test was conducted within 48 hours online on the internet.

Three videos of different content were shown to users. The first one was a presentation with social meaning «The last lecture» by Randy Pausch representing audio-centric information category, the second one was a sport video «Australia Open 2010 Final Federer vs Murray» which is classified as information video-centric category and the last one which was a segment of a popular comedy TV Series «Big Bang Theory» as a representative of narrative-entertainment video category.

Users were instructed to watch the videos, interact any way they wanted with the player and finally, answer the SUS questionnaire, which was also available online with the help of "Google forms". Users watched the three videos in different order to minimize learning effects. Furthermore, after the completion of this procedure, a friendly conversation took place to discover new users' preferences.

### 5.2 Usability Testing Results

SUS score was 79.44 which classifies the application in the middle of good and excellent rating. To sum up, this score means that users are able to use VideoSkip easily and successfully. The users, in general, were able to understand directly its components and seemed to be satisfied. All in all, VideoSkip is an acceptable application.

According to the users the most interesting of the three videos was «The Last Lecture» by Randy Pausch, although it was the one with the fewer users' interactions. On the other hand, in the sport video «Australia Open 2010 Final Federer vs Murray» the greatest number of interactions was observed, even though users commented that it was an indifferent video. Based on users' interactions, we suppose that the more attractive a video is, the less interactions are taking place due to their attentiveness.

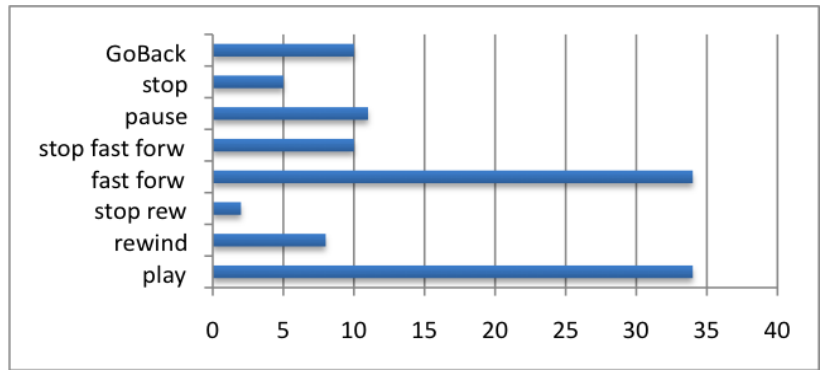

**Fig. 3**. Button usage during the usability test

As fig.3 shows, Play and fast forward buttons were the "popular" ones. Considering users' opinion and the study results, we conclude that some buttons are useless. GoBack button seems to be one of them and consequently did not help us to detect favorite scenes of videos.

One major problem is that the use of player requires high speed internet connection. Some participants were annoyed by the delay of video when the player was in forward state. Moreover, one viewer proposed the construction of a multi-sized web video player.

## 6. CONCLUSION AND ONGOING RESEARCH

Even though we did not test VideoSkip with a large number of subjects, we have found some interesting results. We have developed a user-based video indexing system and we have produced initial evidence that users' interactions with a web video player could play an important role towards enhancing the browsing experience by generating thumbnails. As long as the community of users watching videos online is growing, more and more interactions are going to be gathered and therefore, automatically generated thumbnails would represent effectively the most important scenes of the videos. We also expect that the combination of richer user profiles and content metadata might provide opportunities for additional personalization of the thumbnails.

Based on our hypothesis, we identified the relation between the content of the video and the reactions that occurred. Moreover, we understood the importance of some buttons (play, fast) for browsing a video and the uselessness of others (GoBack).

In current work, we consider the use of only three important buttons: thirty seconds rewind, pause/play and thirty seconds forward, capturing more easily the interactions of the users and (hopefully) producing more representative thumbnails. Additionally, we would like to improve our thumbnail generation algorithm and to compare its effectiveness with other algorithms or with experts' (e.g. video producers') indexing. Moreover, we would like to examine further video contents such as cooking shows or stand up comedy videos through the thumbnail generation process. In future work, we expect that a balanced mix of hybrid algorithms (content-based and user-based) might provide an optimal solution for navigating inside video content on the web.

**Acknowledgments.** This study was partially supported by the European Commission Marie Curie Fellowship program (MC-ERG-2008-230894). We are also grateful to the participants of the study and to many constructive comments by the anonymous reviewers.